\documentclass{ws-jai}
\usepackage[flushleft]{threeparttable}
\usepackage{subcaption}

\begin{document}

\catchline{}{}{}{}{} 

\markboth{Gabriella Montano, Cheukyu~Edward~Tong, Dan Marrone}{A Novel Comb Generator for Frequency Phase Transfer} 

\title{A Novel Comb Generator for Frequency Phase Transfer }

\author{Gabriella Montano$^{1}$, Cheukyu~Edward~Tong$^{2}$, Dan Marrone$^{3}$}

\address{
$^{1}$Center for Astrophysics $-$ Harvard \& Smithsonian, Cambridge, MA 02138, USA, gabriella.montano@cfa.harvard.edu\\
$^{2}$Center for Astrophysics $-$ Harvard \& Smithsonian, Cambridge, MA 02138, USA, etong@cfa.harvard.edu\\
$^{3}$University of Arizona, Tucson, AZ 85721, USA, dmarrone@arizona.edu\\
}

\maketitle



\begin{abstract}
This paper presents the design and testing of a millimeter-wave frequency comb generator developed for the Black Hole Explorer (BHEX) mission, a space Very-Long-Baseline Interferometry (VLBI) mission concept. The heart of BHEX is a dual-band receiver, centered at 90 and 270 GHz. This novel comb generator is based on a microwave phase modulator producing phase-coherent comb signals with multi-octave bandwidth, which will be used to track instrumental delays when injected into the receiver system. The comb generator was tested with astronomical receivers, which confirms its expected operation.

\end{abstract}


\section{Introduction} The Black Hole Explorer (BHEX) is a proposed space mission that seeks to observe black holes using Very-Long-Baseline Interferometry (VLBI). By pairing observations from ground-based radio telescopes with a space telescope, BHEX will achieve an extremely high angular resolution needed to reveal the photon ring surrounding the black hole, as predicted by the Theory of General Relativity \cite{johnson24}.

The receiver system of BHEX has 2 dual-polarization receivers, labeled Rx-L (76-106.6 GHz) and Rx-H (228-320 GHz).  Rx-L is based on a W-band Low Noise Amplifier, while Rx-H is based on the Superconductor-Insulator-Superconductor (SIS) mixer \cite{marrone24}. Tropospheric and instrumental disturbances will affect each receiver, introducing phase fluctuations, which are detrimental to VLBI. To mitigate these effects, we plan to implement the Frequency Phase Transfer (FPT) technique by injecting phase-coherent comb signals into both receivers. This method helps to extend the coherence time of observation of Rx-H by leveraging the higher signal-to-noise ratio (SNR) of Rx-L \cite{rioja23}, ultimately allowing the delay in the receiver signal path to be tracked.

The standard way to generate a microwave comb signal is to pump a Step Recovery Diode (SRD) \cite{friis67} or a Non-Linear Transmission Line (NLTL) \cite{zhang19} with a strong RF signal. For millimeter-wave combs, one can then frequency multiply the microwave comb. Both SRD and NLTL comb generators are commercially available devices, which can potentially generate the required comb; however, they are not optimal for the BHEX instrument for reasons explained here. SRD comb generators produce fast pulses by leveraging the diode's diffusion capacitance for charge storage, resulting in a fast "snap time" \cite{friis67}. The time domain signal produced by an SRD comb generator is sharp, and the ratio of the peak-to-trough of the comb signal depends on the density of teeth in the comb. NLTL comb generators, which are based on a distributed ladder of Schottky varactor diodes, also produce a very sharp time domain signal when pumped with a large input RF signal. These intense amplitude-modulation (AM) based comb signals distort the operation of the SIS mixer. Around the peak of the input pulse, the instantaneous power of the RF signal becomes comparable to the LO power level.  The result is that the SIS mixer behaves as a detector, detecting the envelope of the incident signal rather than down-converting the comb signal, thereby suppressing the heterodyne response. We have performed experiments using an SRD-based comb and found that the spectral output of the SIS mixer is basically the base RF comb independent of the LO frequency, which confirms the direct detection thesis. To avoid such undesirable effect, the injected comb signal must be very weak, which make lab implementation quite challenging.

Owing to the problems caused by AM comb generators, a phase modulation (PM) based comb generator was chosen in our experiments. This paper presents the design and experimental results of this novel comb generator. This comb generator is novel in that it is designed to generate phase-coherent combs for injection into the 1 mm and 3 mm receivers simultaneously. The designed comb generator also covers a very wide band and the design goal is for a space-qualified module deployable for the BHEX mission.  We report here the initial lab validation tests of the comb using the next-generation Event Horizon Telescope (ngEHT) 86/115 GHz HEMT \cite{tong24a} and the wideband Submillimeter Array (wSMA) \cite{grimes17} receivers. We also present the results of the tests conducted at the Kitt Peak Radio Observatory, Arizona, using a dual-band receiver system.

\section{Specifications and Design Approach}

The comb generator is expected to cover the BHEX operating band of Rx-H for 4 different LO tunings. Since the IF is 4-12 GHz, the comb is expected to cover 12 GHz on either side of the LO frequency, a span of 24 GHz. The output comb should also cover the operating frequency band of Rx-L, which is one-third that of Rx-H. The adopted design methodology is to generate a phase-modulated comb around a microwave carrier, at a frequency $F_0$, which is phase modulated by a low frequency modulating signal, $F_{mod}$. A schematic of the comb generator is given in Fig. 1. A microwave signal source, with frequency, $F_0\sim10$ GHz, drives the local oscillator (LO) port of a double-balanced mixer. 


\begin{figure}[!t]
    \centering
    \includegraphics[width=1\linewidth]{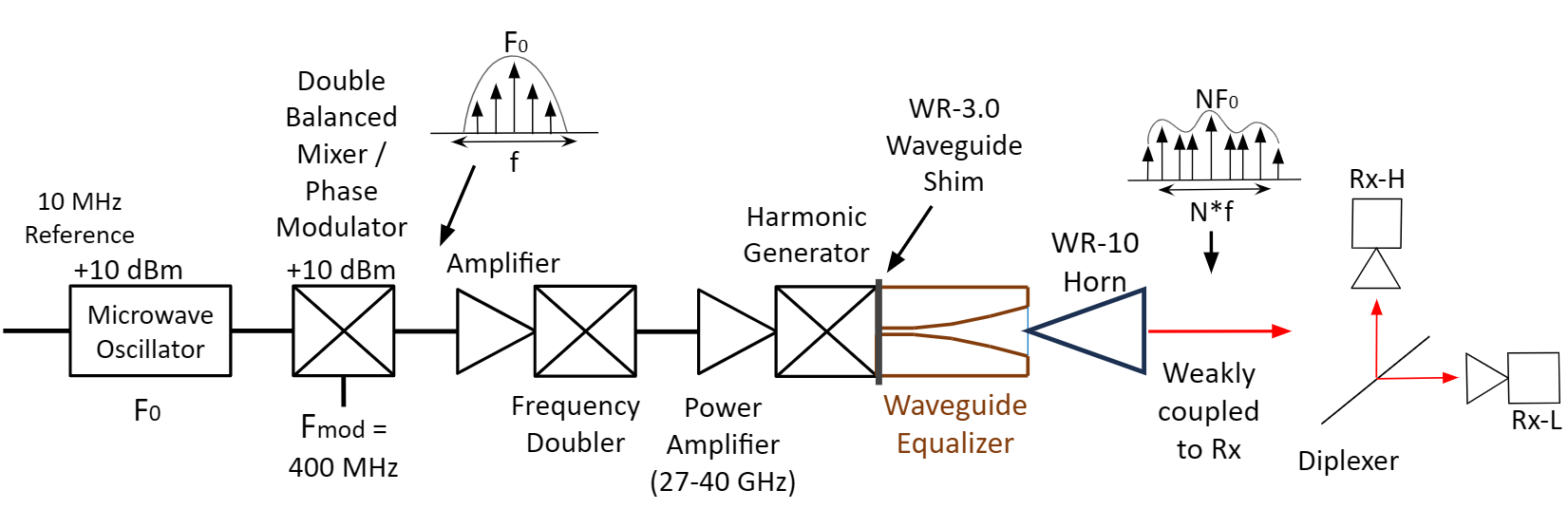}
    \caption{A block diagram of the comb generator setup, The radiated comb signal is sent to the receiver pair Rx-L (90 GHz) and Rx-H (270 GHz) through an optical diplexer.}
    \label{fig:enter-label}
\end{figure}

The heart of the phase modulation scheme is a double-balanced-mixer.
A 400 MHz signal, tied to the same 10 MHz reference as the carrier, is connected to the mixer's intermediate frequency (IF) port, while the carrier pumps the mixer via its LO port. 
The phase modulator output, taken from the RF port of the mixer, is a narrow base comb, in which the carrier is flanked by a few sidebands spaced at 400 MHz apart. This base comb is then amplified to drive the harmonic generator, which is a WR-10 wideband frequency multiplier with a coaxial input port \cite{tong25}. It is specified to work between 75 and 110 GHz with a flat frequency response. Several multipliers have been tested to produce an optimal output from the harmonic generator. The best results were obtained from a model W3 tripler from Pacific Millimeter Products. This multiplier was chosen over other candidates, due to its wideband operation as well as its uniform and stable comb generation. 
%

An important criterion for millimeter-wave combs is that the output comb teeth should be uniform, with few missing teeth and few low-amplitude teeth. The W3 multiplier performs well in this area, even up to 300 GHz. Additionally, the W3 has space heritage. The output of the harmonic generator is connected to a waveguide equalizer, which is basically a waveguide shim, followed by a waveguide taper, before being radiated through a WR-10 horn. Another advantage of the W3 multiplier is that the location of its internal Schottky diode is physically close to the waveguide flange. This is important because it reduces the occurrence of resonances in the frequency response of the comb. The equalizer promotes the higher order combs required for Rx-H, while attenuating the strong lower order combs, needed for Rx-L. Without the equalizer, the comb teeth for Rx-L will be many 10s of dB above those for Rx-H.

The comb signals will be injected into the BHEX receiver beam waveguide through a small opening in the calibration load, which is common to both receivers. Therefore, the comb signal will only be active during a specific calibration cycle, in which each receiver will be weakly coupled to the comb generator through an optical diplexer \cite{carter24} which splits the receiver beam to feed both Rx-H and Rx-L.

\section{Phase Modulation and the Modulation Index} An ideal phase modulated signal can be described by the following equation:
\begin{equation}
    X(t)=A_c\cos[\omega_ct+\beta cos(\omega_mt)],
\end{equation}
where $A_c$ is the carrier signal amplitude, $\omega _c$ and $\omega _m$ are the carrier and modulation frequencies, respectively, and $\beta$ is the modulation index \cite{stremler90}.

To understand the relationship between $\beta$ and the amplitude of the modulation signal in our setup, we tracked the phase of a 2 GHz carrier modulated by a 0.1 Hz sine wave using two different modulation amplitudes. The results are displayed in Fig. 2. When the modulating signal amplitude was weak ($0.6V_{pp}$ in the top figure), the carrier phase varied sinusoidally with an amplitude of 2 radians peak-peak, corresponding to $\beta\sim 1$ according to Eq. (1), where the phase is described by $\phi=\beta \cos(\omega_mt)$. With a stronger modulation ($1.6V_{pp}$), the measured phase was a clipped sine wave, and $\beta$ was saturated. The Fourier transform of this phase waveform yields an amplitude of $\sim 1.5$. Thus, for the double balanced mixer we used, $\beta_{\rm max}\sim$ 1.5. This value cannot be increased further due to the aforementioned saturation effects, and this demonstrates the limitations of the double balanced mixer as a phase modulator.
\begin{figure}[!t]
    \centering
    \includegraphics[width=4in]{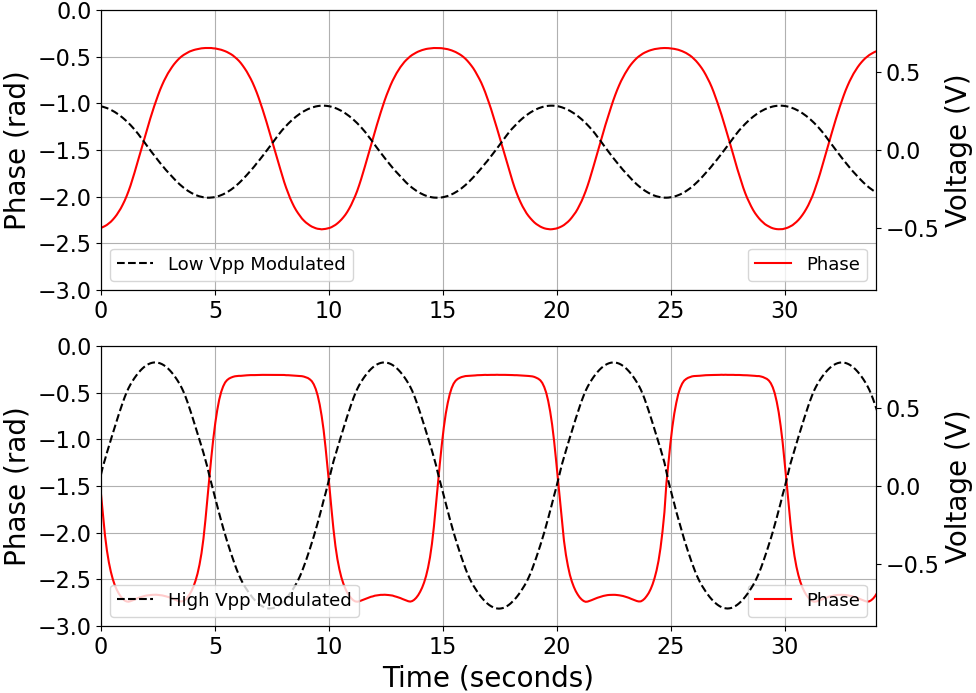}
    \caption{Measured phase change of a 2 GHz carrier when a 0.1 Hz sine wave (black dotted curve) was applied to the phase modulator for 2 different modulating signal amplitudes -- $0.6V_{pp}$ and $1.6V_{pp}$.}
    \label{fig:pure-sine}

\end{figure}

According to Carson's Rule \cite{pieper01}, the number of significant sidebands in phase or frequency modulation is given by:

\begin{equation}
    N_S = 2\beta+1,
\end{equation}

This rule of thumb suggests that our phase modulator can only generate a narrow comb because $\beta$ is small. However, the microwave comb feeds a harmonic generator to generate the desired millimeter-wave comb. Let $n$ be the frequency multiplication used. The $n$-th order comb emerging from the harmonic generator, $X_n(t)$ can be derived from Eq. (1) and can be written as follows:

\begin{equation}
    X_n(t) = A^{(n)}_c \cos[n\omega_ct+n\beta cos(\omega_mt)],
\end{equation}

where $A^{(n)}_c$ is the amplitude of the multiplied comb. This equation shows that the modulation index of the $n$-th order comb is $n\beta$. It follows that the number of significant sidebands, or the effective width of the comb is increased by a factor of $n$. Note that the percentage bandwidth covered by one order of the comb is unchanged from the base comb.

\begin{figure}[!t]
    \centering
    \includegraphics[width=4in]{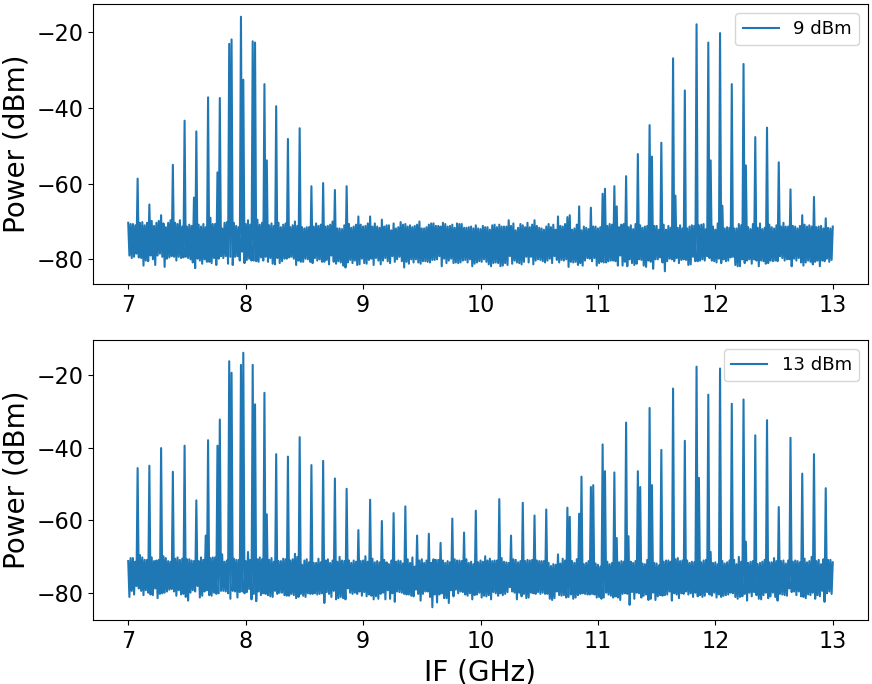}
    \caption{Second and Third order combs observed when a 4 GHz carrier was modulated by a 100 MHz signal with applied power of 9 dBm (Top) and 13 dBm (Bottom).}
    \label{fig:overlaporders}

\end{figure}

We have observed the spectra of the base comb as well as the different orders of the frequency-multiplied comb using a spectrum analyzer. We have found that the base comb is minimally changed beyond an incident modulation power of +13 dBm. This agrees with the lower plot in Fig. 2 which shows that $\beta$ is saturated at this level of modulation. Another effect of strong modulation is that 
different orders of the combs may overlap with each other, forming a fuller coverage of the spectrum. This is illustrated in Fig. 3. In fact, the frequency of each observed comb tooth, $F_{obs}$, bears the following relationship with $F_0$ and $F_{mod}$,
\begin{equation}
\label{eq:order}
    F_{obs}=(nF_{0})\pm(mF_{mod}), 
\end{equation}

where $n$ is the comb order, and $m$ is the number of teeth from the center frequency in either direction; $m$ can be positive or negative. This equation shows that if $F_0$ is an integer multiple of $F_{mod}$, then any comb tooth may be produced by a combination of $n$ and $m$ under strong modulation. The limitations for this theory are governed by the Bessel functions, where the $m$-th tooth is given by the Bessel function of the first kind, $J_m(\beta)$ \cite{bowman58}, which has a series expansion
\begin{equation}
    J_m(\beta)=\sum_{k=0}^{\infty}\frac{(-1)^k}{k!\Gamma(k+m+1)}(\frac{\beta}{2})^{2k+m},
\end{equation}
where as $m$ approaches infinity, the amplitudes of $J_m(\beta)$ decrease. Additionally, according to Eq. (2), for $m>>\beta$, the amplitude of the comb teeth will become increasingly small. Both of these factors impose a maximum usable value for $m$, and so signals of infinite frequency cannot be generated from Eq. (4). As such, for a given comb tooth, the greatest contribution comes from the combination where $m$ is the minimum. Nevertheless, this property is useful for back-filling the nulls in a phase modulation spectrum with adjacent comb order. Such an effect is expected to be more pronounced for the higher-order multiplied combs at millimeter wavelengths. 

The overall strength of the $n$-th order comb is harmonic generator dependent. For the harmonic generator used in our experiment, the Schottky diode sits in front of a waveguide backshort. For backshort distances close to odd multiples of $\lambda_g/4$, where $\lambda_g$ is the guided wavelength inside the harmonic generator, strong outputs are expected. In our case, we the expected peak outputs are around 90 and 270 GHz for our WR-10 harmonic generator.

\section{Comb Production for BHEX Receiver}
In order for a robust spectrum to be produced, the output of the harmonic generator requires optimization. One important goal for our design is that the comb teeth provide good coverage throughout frequency band for both Rx-L and Rx-H with minimal missing tones. Since the harmonic generator is expected to produce a very strong output for Rx-L, we introduce a waveguide high-pass filter (HPF), which acts more as a waveguide equalizer, at the output of the W3 frequency multiplier. Without a waveguide equalizer, the power output from the frequency multiplier is on the order of $\sim$mW, which is more power than is required. The optimal HPF-waveguide equalizer configuration was determined experimentally: a 0.254 mm thick WR-3.0 shim followed by a WR-6.5-to-WR-10 waveguide taper. Electromagnetic simulation shows that this HPF incurs a 70 dB attenuation of an output signal at 80 GHz from the WR-10 multiplier. In practice, the WR-3.0 waveguide shim creates a "cavity" together with the backshort of the multiplier, which enhances the harmonic output in the 240-320 GHz band.

\begin{figure}[!t]
    \centering
    \includegraphics[width=7in]{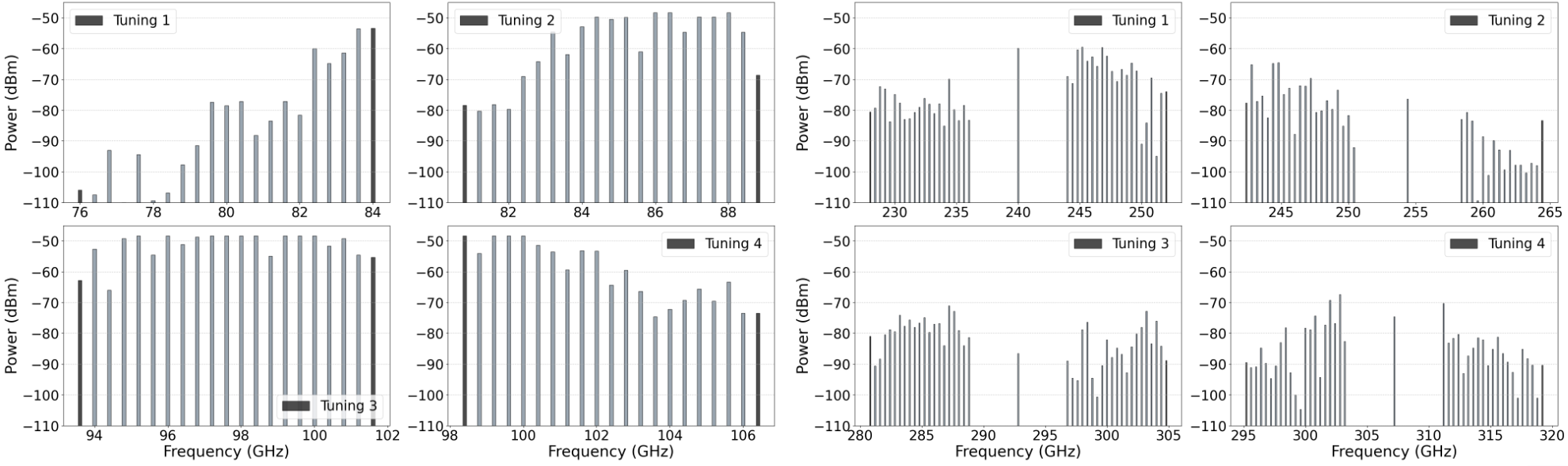}
    \caption{Composite spectra measured with a spectrum analyzer equipped with a WR-3.4 harmonic mixer depicting the comb coverage for Rx-L (left) and Rx-H (right). The power levels are only approximate because we do not have a calibration for the harmonic mixer. Also, we are only measuring the comb teeth falling in the 4-12 GHz IF for each LO tuning.}
    \label{fig:composite-spectra}

\end{figure}

Fig. 4 compiles the output of the comb generator, as measured by a spectrum analyzer, equipped with a WR-3.4 harmonic mixer operating with a harmonic number of 24. We have also chosen the carrier frequency as 1/24 that of the LO frequency of Rx-H. For all 4 LO tunings of Rx-H, a decent comb with relatively high power is observed at a teeth spacing of 400 MHz, the modulating frequency. Ample comb power is also produced for all 4 corresponding tunings of Rx-L, even with the strong attenuation presented by the waveguide equalizer. Some amplitude fluctuations are observed in the comb spectra for Rx-H. This is likely caused by the interaction between the waveguide shim and the harmonic generator, which introduces harmonic resonances in the high band. Additionally, the harmonic mixer used in conjunction with the spectrum analyzer operated at a high harmonic number such that its conversion loss as a function of frequency is not expected to be uniform because many idler frequencies are involved and the impedances at these idler frequencies are uncontrolled and varied.

\begin{figure}[!t]
    \centering
    \includegraphics[width=3in]{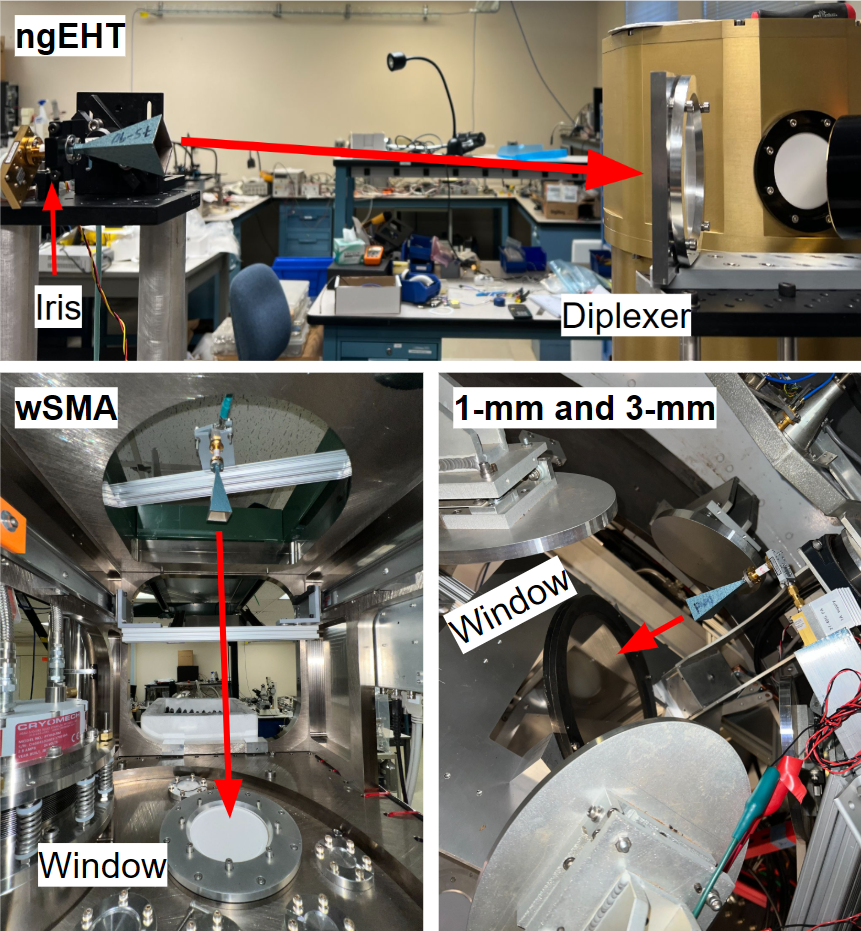}
    \caption{Comb injection test setup using the ngEHT 86/115 GHz receiver (Top), wSMA receiver (Bottom Left), and 1mm and 3mm receiver (Bottom Right). For ngEHT receiver, the comb and horn are mounted on a plate, directed towards the back side of the optical diplexer, which acts as a $\sim$20 dB coupler for the comb signal. For wSMA receiver, the horn is mounted above, pointing downward toward the window. For the Kitt Peak receivers, the comb generator is mounted on the side, with the horn pointing towards a wire grid polarizing diplexer placed in front of the vacuum windows of the receivers}
    \label{fig: wSMA}

\end{figure} 
\section{Experimental Results with Astronomical Receivers}

We have tested the comb generator with three different receivers: the ngEHT 86/115 GHz Side-car receiver \cite{tong24a}, the wSMA receiver \cite{grimes17}; and the four-band receiver installed in the 12-meter ALMA prototype antenna on Kitt Peak, AZ \cite{lauria21}. The setup for each test is shown in Fig. 5. The ngEHT and wSMA receivers were configured as stand-alone in-lab receivers, and therefore we could only use them to check one part of the output spectrum of the comb generator for each receiver.

\subsection{ngEHT Receiver}
This dual-polarization receiver is based on a wideband cryogenic low-noise amplifier. It is equipped with an optical diplexer, which separates the incident 85-115 GHz signal beam from the higher frequency signal beam, by reflection. Referring to 
Fig.~\ref{fig: wSMA}, we positioned the comb generator on the backside of the diplexer such that the frequency comb around 90 GHz was weakly coupled to the receiver. The E-plane of the comb generator horn was rotated by $45^o$ so that both polarization channels of the receiver could pick up the comb signal. This receiver operates with a Single-Side-Band (SSB) down-converter. Fig.~\ref{fig: spectra2} shows a comb spectrum extending between 81 and 88 GHz, with an SNR of 10 dB at a resolution bandwidth of 3 MHz.


\subsection{wSMA Receiver}

We also performed a test with the wSMA receiver, which is equipped with Double-Side-Band (DSB) SIS mixers operating in the 200 and 300 GHz range. For this measurement, the comb generator was mounted on slats directly above the vacuum window of the receiver. The E-plane of the comb generator horn was oriented to allow coupling to both polarizations of the receiver. At the location where the horn was mounted, the receiver beam size was much larger than that of the horn aperture. Therefore, the coupling between the receiver and the comb generator is expected to be weak ($<\sim 1\%$). Absorbers were installed around the horn aperture to reduce spurious reflection.

\begin{figure}[!t]
    \centering
    \includegraphics[width=4in]{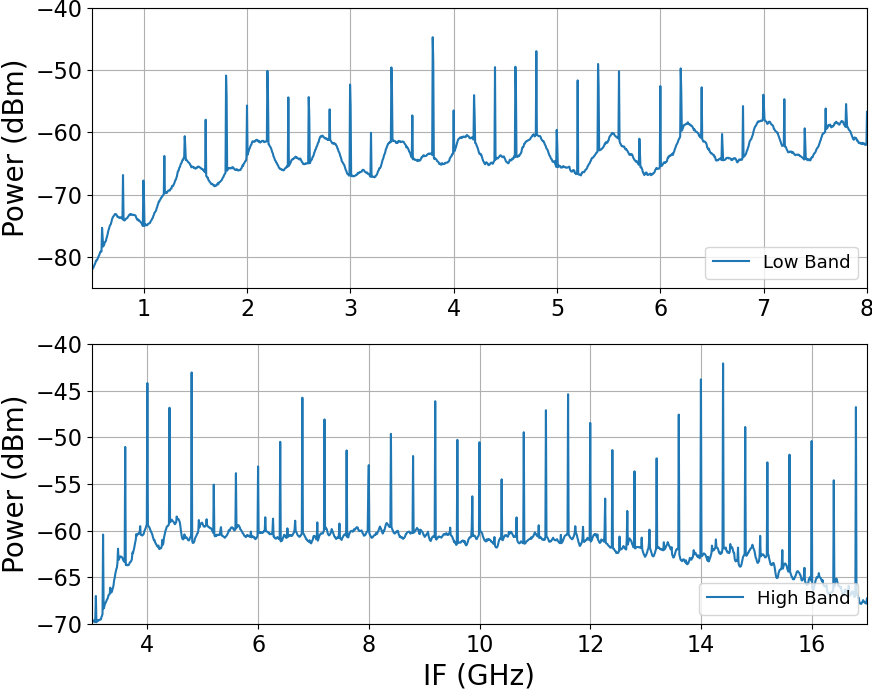}
    \caption{IF output spectra recorded by a spectrum analyzer (RBW 10 kHz) connected to the output of ngEHT 86/115 GHz receiver (Top) and wSMA 240 GHz receiver (Bottom) during the comb injection experiment. }
    \label{fig: spectra2}

\end{figure}

The wSMA receiver was tuned to multiple frequencies to cover the intended pass band of BHEX. At each LO tuning, the base frequency of the comb, $F_0$, was set to be 1/24 times the LO frequency. Thus the receiver is expected to couple the 24th order comb as specified by (\ref{eq:order}). The spectrum shown in the bottom plot of Fig.~\ref{fig: spectra2} was recorded at an LO frequency of 240 GHz. Observed comb teeth demonstrated a strong SNR, generally $>$20 dB, covering the 4 - 12 GHz IF band. Since this is a DSB receiver, the comb signal extends from 228 to 252 GHz.

Compared to the low-band comb experiment, the SNR is stronger. This can be explained in a number of ways: beam coupling differences; the wSMA receiver operates in DSB mode, which causes 2 comb teeth to overlap at the output; also, the low frequency cutoff from the WR-3.0 shim may be a bit too severe. In the final BHEX design, further considerations of beam coupling and the required amount low frequency equalization will have to be defined.

\subsection{Arizona Radio Observatory 12m Telescope}

\begin{figure}[!t]
    \centering
    
    \begin{subfigure}[t]{0.48\textwidth}
        \centering
        \includegraphics[width=\linewidth]{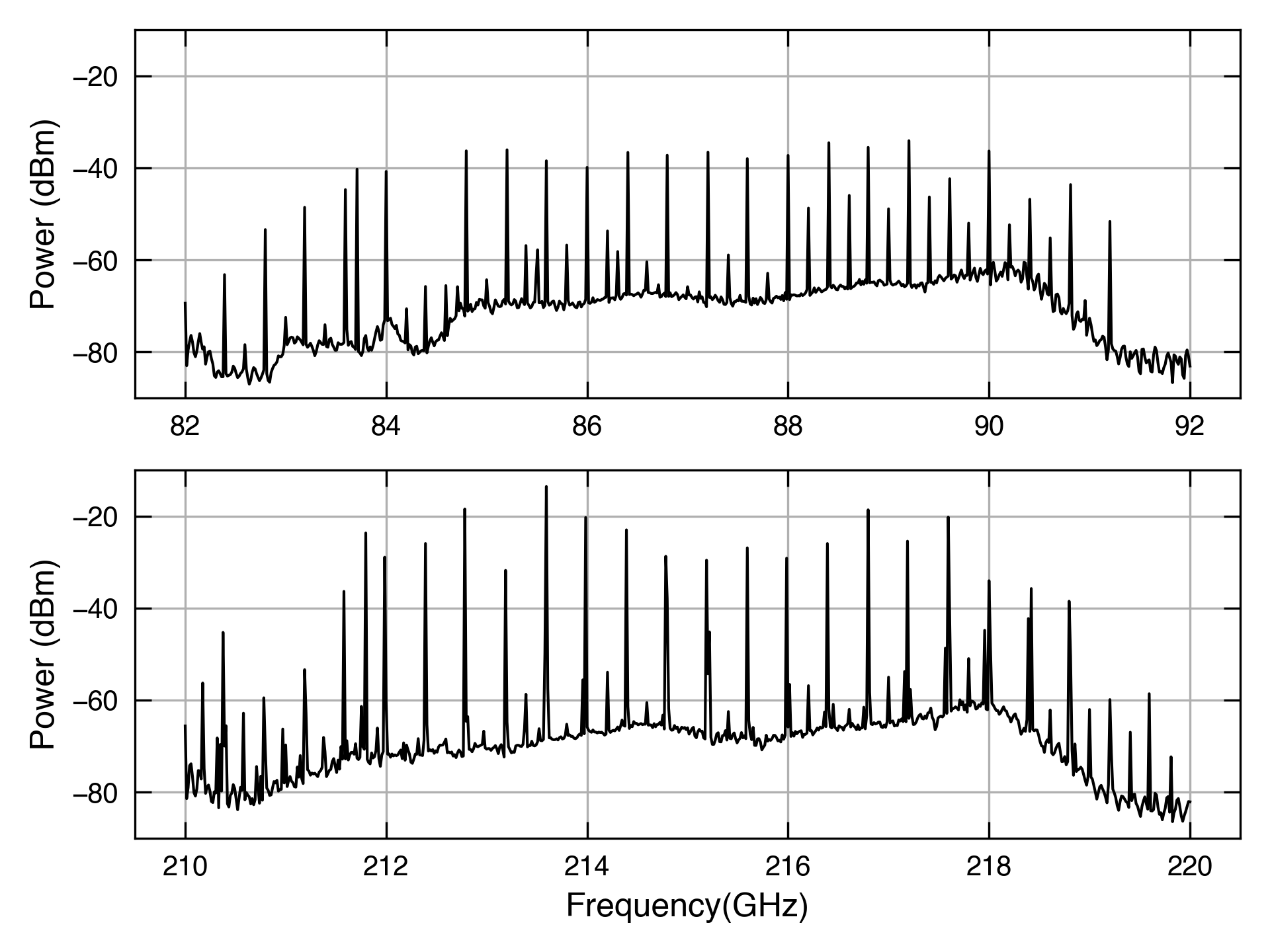}

        \label{fig:1mm3mm}
        
    \end{subfigure}
    \hfill
    \begin{subfigure}[t]{0.48\textwidth}
        \centering
        \includegraphics[width=\linewidth]{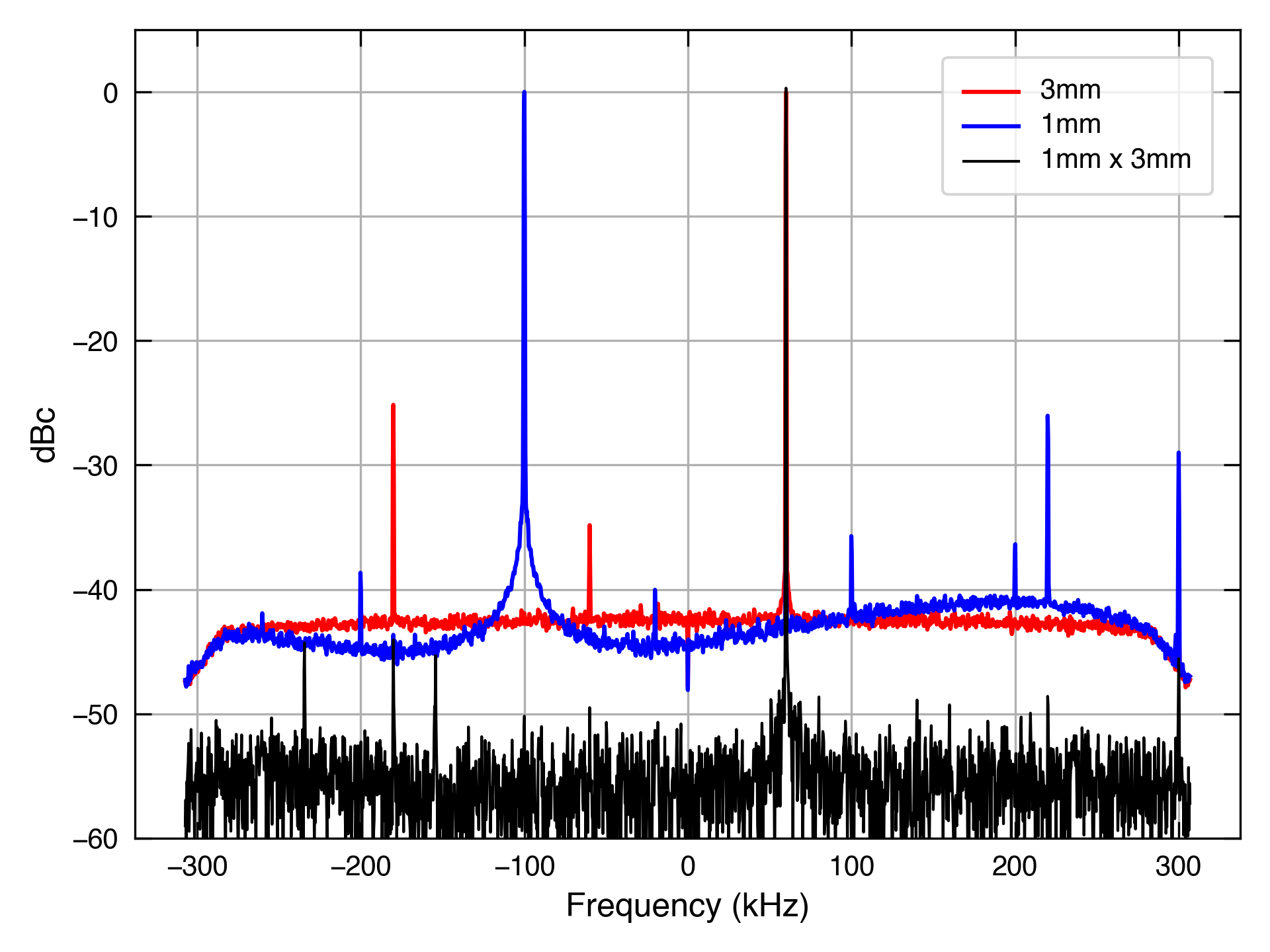}
        \label{fig:tone_xpspec}
    \end{subfigure}

    \caption{
    Left: Tone generator output spectra as measured at 10 kHz resolution through the Kitt Peak 12m telescope 4-band receiver in the 3mm (Top) and 1mm (Bottom) bands. These spectra correspond to an IF range of 1-11~GHz, with the frequency axis reversed because both are measured in the receiver lower sideband.
    Right: Narrow band spectra of one tone in each band, with the cross-correlation spectrum shown in black after applying a frequency shift to correct the intentional 160~kHz frequency offset. The tones are normalized to their peak amplitudes, and the cross-spectrum is normalized to the peaks of the 1 mm and 3 mm spectra.}
    \label{fig:1mm3mmspec}

\end{figure}

The experiment conducted at the 12 m Telescope was significant in that it allowed the comb generator to be tested simultaneously in the 3 mm and 1 mm bands. The four-band receiver normally observes with only one band at a time, sampling both polarizations. For this test, additional mirrors and a wire grid were inserted to direct the sky signal (in this case, the comb generator) to one polarization in each of the 3 mm and 1 mm receiver bands. The IF signal path was rearranged to deliver the Lower-Side-Band (LSB) of each receiver to subsequent amplification and down-conversion hardware. The LOs were tuned to 93.00000~GHz and 221.00016~GHz, respectively, and locked to a common 10~MHz reference derived from the site's active hydrogen maser. The 1-11~GHz IF spectrum for each band is shown in Fig.~\ref{fig:1mm3mmspec}, with the measured IF frequencies converted back to their corresponding RF frequencies. The nominal IF band of the receivers is 4-8~GHz, determined by the frequency coverage of the IF amplifiers, but the tone clearly shows up across a wider range. A prominent comb spectrum is visible in both bands, averaging $\sim 35$ dB of SNR in a 10 kHz channel. 

To verify the utility of the comb generator as a tool for measuring the time delay between receiver signal paths, we recorded simultaneous IF output signals in both bands for cross correlation. 
We selected tones at 87.20000~GHz and 215.20000~GHz, 
which are down-converted to 5.80000 and 5.80016~GHz in the 3 mm and 1 mm IFs, respectively. These were mixed with a 7 GHz DRO, also phase locked to the maser, placing them at 1.20000 and 1.1984~GHz, where they could be digitized by a dual-channel software defined radio (SDR), an Analog Devices ADALM-PLUTO. 
Data were captured with a sampling rate of 300~kHz. 
The SDR spectra are shown in Fig.~\ref{fig:1mm3mmspec} (right). The 160kHz offset between the IF frequencies was introduced to ensure that cross-talk between the channels was not a significant source of coherence, and the data were phase rotated to correct the frequency difference before cross correlation. The cross correlation shows nearly perfect coherence, approximately 99.8-99.9\%, with the small variation introduced by reducing the amplitude of the 12~GHz drive by 2 dB. Path length changes were also observed in the complex phase of the cross-correlation peak when additional lengths of cable were added to one receiver band. This site test validates the comb generator's performance and overall suitability for dual-band phase tracking and calibration.

\section{Outlook}
This paper presents the design and experimental results of a novel millimeter-wave comb generator based on a double-balanced mixer functioning as a phase modulator. Paired with the FPT technique, the comb generator has target applications for high precision VLBI observations, such as ngEHT and BHEX. When tested over the BHEX frequency ranges with astronomical receivers, we have observed output combs with good signal-to-noise ratio in addition to a fairly uniform tone coverage over the receiver pass-band. Additionally, phase tests conducted with the 1mm and 3mm receivers at Kitt Peak Arizona Radio Observatory validate its functionality as a dual-band phase tracker, as it maintained stable phase measurements across both bands through path length changes. These results demonstrate that this comb generator is ready to be deployed for receiver phase tracking as required by the FPT algorithm.

\noindent





\bibliographystyle{ws-jai}
\bibliography{references}


\end{document}